# Coherent-Incoherent Crossover of Charge and Neutral Mode Transport as Evidence for the Disorder-Dominated Fractional Edge Phase

Masayuki Hashisaka,[1,2]* Takuya Ito,[3] Takafumi Akiho,[1] Satoshi Sasaki,[1] Norio Kumada,[1] Naokazu Shibata,[3] Koji Muraki[1]

[1]NTT Basic Research Laboratories, NTT Corporation, 3-1 Morinosato-Wakamiya, Atsugi, Kanagawa 243-0198, Japan.

[2]JST, PRESTO, 4-1-8 Honcho, Kawaguchi, Saitama 332-0012, Japan.

[3]Department of Physics, Tohoku University, Sendai, Miyagi 980-8578, Japan.

*To whom correspondence should be addressed; Email: masayuki.hashisaka.wf@hco.ntt.co.jp

Couplings between topological edge channels open electronic phases possessing nontrivial eigenmodes far beyond the noninteracting-edge picture. However, inelastic scatterings mask the eigenmodes' inherent features, often preventing us from identifying the phases, as is the case for the quintessential Landau-level filling factor $v = 2/3$ edge composed of the counter-propagating $v = 1/3$ and 1 (1/3-1) channels. Here, we study the coherent-incoherent crossover of the 1/3-1 channels by tuning the channel length in-situ using a new device architecture comprising a junction of $v = 1/3$ and 1 systems, the particle-hole conjugate of the 2/3 edge. We successfully observed the concurrence of the fluctuating electrical conductance and the quantized thermal conductance in the crossover regime, the definitive hallmark of the eigenmodes in the disorder-dominated edge phase left experimentally unverified.

**I. INTRODUCTION**

Manipulating edge excitations is the key to exploiting exotic quasiparticles in topological matter for future information technologies. Coulomb interaction plays a critical role not only in determining the bulk ground state but also in shaping the edge excitations. Its role becomes pivotal when more than one edge channel is involved. The inter-channel interaction can drive the edge to a nontrivial electronic phase, transforming the elementary edge excitations (eigenmodes) stemming from the bulk topological order into entirely different modes. Among various cases [1-22], downstream charge and upstream charge-neutral (neutral) modes in the fractional quantum Hall (FQH) state at Landau-level filling factor $v = 2/3$, the hole-conjugate of the $v = 1/3$ state, embody the essence of the interacting edge phases in its entirety [1-3].

The interplay of electron correlation and confinement potential determines the electron-density profile $n_e(x)$ of a two-dimensional (2D) electron system (2DES) near the sample edge [23]. At $v = 2/3$ ($= 1 - 1/3$), a $v = 1$ strip forms along the edge, accompanied by counter-propagating channels (hereafter "1/3-1 channels") reflecting the filling-factor discontinuities $\delta v = -1/3$ and $1$ on either side [24-30]. Edge dynamics at $v = 2/3$ exhibit an intriguing phase diagram [Fig. 1(a)] governed by Coulomb interaction and tunneling through random impurities ("disorder," in other words) between the 1/3-1 channels. At zero temperature, the renormalization of interaction by disorder drives the system to follow different fates depending on the bare interaction strength represented by a dimensionless parameter $\Delta$ ($\geq 1$) [1-3]. When the interaction is weak [weak-coupling phase, blue region in Fig. 1(a)], the system holds upstream and downstream eigenmodes, both charged, that reflect the renormalized interaction strength ($\Delta_r > 3/2$) [31]. In contrast, when the bare interaction is strong (strong-coupling phase, or "disorder-dominated phase [1, 2]," red region), the system is driven toward the strong-interaction limit ($\Delta_r = 1$), referred to as the strong-disorder fixed point, and thus the eigenmodes inherent to the limit, the charge and neutral modes, emerge irrelevantly to the bare interaction strength.

In a real system of a finite length and at a finite temperature, the eigenmodes differ slightly from those exactly at the fixed point; yet, at sufficiently low temperatures, the system stays in the basin of attraction of the fixed point ($\Delta_r \cong 1$), and the charge and neutral modes well explain the transport phenomena. However, despite the essential difference with the weak-coupling phase, unambiguously identifying the strong-coupling phase is an experimental challenge. In most of the transport regimes distinguished by the channel length $L$, the strongly-interacting 1/3-1 channels show similar electrical and thermal conductance to those in the weak-interaction case [Fig. 1(b)] [3]. The "coherent" and "incoherent" regimes exhibit quantized electrical conductance $4G_e/3$ ($G_e = e^2/h$ with $e$ elementary charge and $h$ Planck's constant) and $2G_e/3$, respectively, in both strong- and weak-coupling phases. Interestingly, theory predicts a characteristic crossover between the coherent and incoherent regimes ("mesoscopic" regime) only in the strong-coupling phase, which would present the unique hallmark of the charge and neutral modes near the fixed point, that is, the concurrence of fluctuating electrical conductance [3,32,33] and quantized thermal conductance $G_Q = \pi^2 k_B^2 T_e/3h$ (with $k_B$ Boltzmann's constant and $T_e$ electron temperature) [3]. So far, most previous experiments investigated the neutral mode only in the incoherent regime with electrical conductance $2G_e/3$ [7-9, 11-13, 16-21]. The mesoscopic regime remains unexplored due to the difficulty in fabricating short channels below the inelastic scattering length in a natural $v = 2/3$ system [34, 35]. To overcome this problem, we employ a device comprising a junction of $v = 1/3$ and 1 systems (1/3-1 junction), the particle-hole conjugate of the 2/3 edge, which enables us to fine-tune the 1/3-1 channel length below the inelastic scattering length.

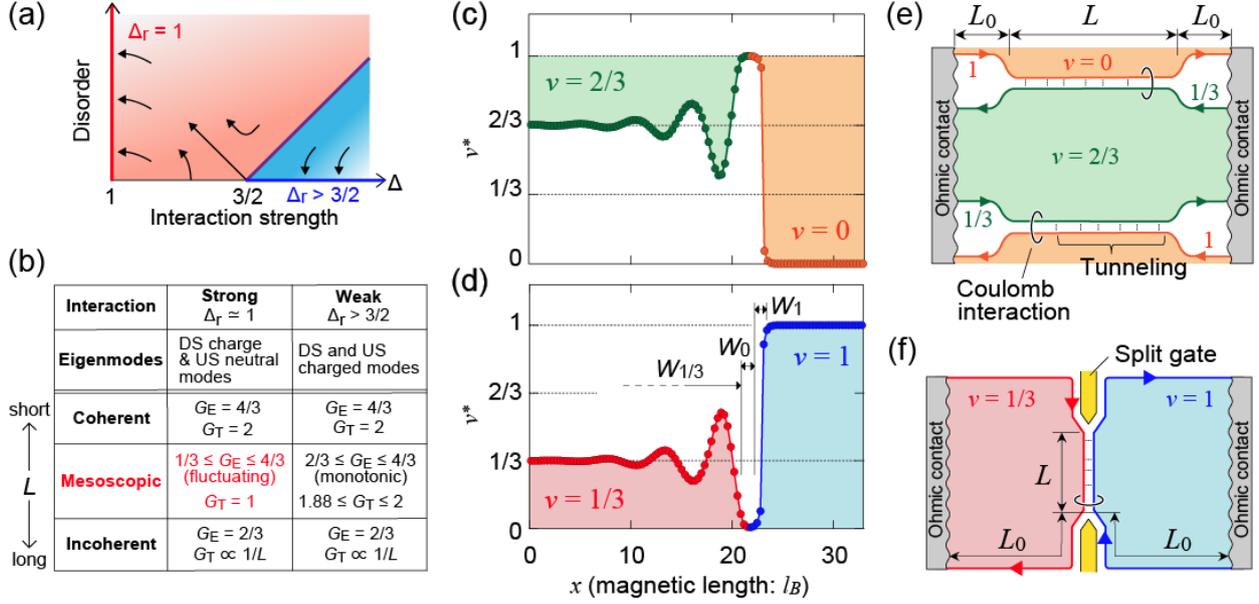

FIG. 1. (a) Phase diagram of the 2/3 edge at zero temperature [1]. Red (blue) region is the strong-coupling (weak-coupling) regime. (b) Summary of electrical ($G_E$) and thermal ($G_T$) conductance of conventional $v = 2/3$ device in units of $G_e$ and $G_Q$, respectively [3]. DS (US) stands for "downstream (upstream)." (c) DMRG calculation results of $n_e(x)$ profiles across the 2/3-0 and (d) 1/3-1 junctions, plotted in units of filling factor $v^*$. Both profiles show the counter-propagating 1/3-1 channels, demonstrating the particle-hole symmetry between the junctions. The profile in (d) enables us to estimate the compressible $v = 1/3$ ($W_{1/3}$) and $v = 1$ ($W_1$) strip widths and their separation ($W_0$). We calculated channel capacitances using the 2D finite element method and obtained $\Delta \simeq 1.2$, which indicates that our 1/3-1 channels are in the strong-coupling phase (see Appendix B). (e) Schematic of conventional $v = 2/3$ device and (f) our 1/3-1 junction. Interacting 1/3-1 channels (length $L$) are connected to ohmic contacts via noninteracting leads (length $L_0$).

## II. RESULTS

### A. 2/3 edge versus 1/3-1 junction

To gain insight into the microscopic structure of the 1/3-1 junction and thereby examine its particle-hole symmetry with the 2/3 edge, we performed density-matrix-renormalization-group (DMRG) calculations for its electron-density profile $n_e(x)$ [36]. Figures 1(c) and (d) are the obtained profiles transverse to a 2/3 edge (2/3-0 junction) and a 1/3-1 junction, respectively, with the vertical axis $v^*(x) = n_e(x)/n_\varphi$ with $n_\varphi$ the density of flux quanta. For both calculations, we included an electrostatic potential gently varying with $x$ to impose a density difference between the left- and right-hand sides (see Appendix A). In Fig. 1(d), $v^*(x)$ approaches $v^*(x) \cong 1/3$ and 1 on the left- and right-hand side, respectively, indicating that incompressible quantum Hall (QH) states develop in the bulk regions. Notably, a narrow $v^*(x) = 0$ region emerges at the boundary, with $v^*(x)$ oscillating on the $v = 1/3$ side in the same way as in an isolated $v = 1/3$ system [36]. Comparing $n_e(x)$ in Figs. 1(d) and (c), we see that a particle-hole symmetry, $v^*(x) \leftrightarrow 1 - v^*(x)$, holds between the 1/3-1 junction and the 2/3 edge. We confirmed that the microscopic $n_e(x)$ structure in Figure 1(d) [(c)] is robust against small changes in the potential due to the adjustability of the position of the $v^* = 0$ (1) strip.

The DMRG calculations also enable us to estimate $\Delta$. From the widths ($W_{1/3}$ and $W_1$) and separation ($W_0$) of the compressible strips [Fig. 1(d)], we calculated the relevant capacitances using a 2D finite-element method [15, 31, 37, 38]. The obtained capacitance values give $\Delta \cong 1.2$, which ensures that our 1/3-1 channels reside in the strong-interaction regime, where disorder drives the system toward the strong-disorder fixed point (see Appendix B).

As we explain below, compared to the 2/3 edge, our 1/3-1 junction is advantageous for accessing the mesoscopic regime. Figures 1(e) and (f) compare the edge-channel configurations in a conventional $v = 2/3$ device and our 1/3-1-junction device. The mesoscopic regime takes place when the condition

$$l \ll \min(L_0, L_T) \ll L \ll L_{in} \qquad (1)$$

is satisfied [3]. Here, $L$ is the length of the interacting 1/3-1 channels, $L_0$ is that of the noninteracting channels serving as leads connecting the interacting region with the ohmic contacts, $l$ is the characteristic length of disorder, $L_T \sim hv_\sigma/k_B T_e$ (with $v_\sigma$ neutral-mode velocity) is the thermal length, and $L_{in}$ is the characteristic length of inelastic inter-mode scattering. Condition (1) means that, to investigate the mesoscopic regime, $L$ must be long enough to cause disorder-induced inter-mode scattering while short enough to suppress inelastic processes causing energy dissipation. In the conventional device [Fig. 1(e)], transition regions unavoidably exist near the ohmic contacts, where the 1/3 and 1 channels no longer interact, playing the role of leads (of unknown length $L_0$). As these regions are short, the condition $l \ll \min(L_0, L_T)$ is not generally met. Most importantly, the channel length $L$ is fixed by design, with no chance to examine the $L$ dependence. In contrast, our 1/3-1 junction [Fig. 1(f)] is electrostatically defined, where $L$ is tunable in situ by a split gate. The 1/3 and 1 channels are noninteracting outside the junction region, serving as leads of well-defined length $L_0$. This guarantees $L_0 > L_T$ even at low temperatures, which makes it possible to examine the $L$ dependence under the condition $l < L_T < L < L_{in}$. In addition, by measuring the electrical conductance $G$ across the junction, the electrical conductance of the 1/3-1 channels is directly obtained as $G_e - G$ without its being affected by the unknown bulk conduction. We also note that the full spin polarization of $v = 1/3$ and 1 removes unnecessary complications due to the spin degree of freedom that might come into play at $v = 2/3$ [39].

**B. Electrical-conductance and currnet-noise measurements**

Figure 2(a) shows a schematic of our sample and experimental setup. The filling factors in the left (red) and right (blue) gated regions [Fig. 2(b)] are set to $v = 1/3$ and 1, respectively, at the perpendicular magnetic field $B = 10$ T. Applying a large negative voltage $V_S$ to the split-gate electrode (yellow, aperture size $L_g$) located between the left and right gates depletes the 2DES underneath and forms a narrow junction, along which $v = 1/3$ and 1 channels counterpropagate. When the 1/3-1 channels are in the strong-coupling phase, the charge and neutral modes would propagate from the downside to the upside and from the upside to the downside, respectively. We measure the differential conductance $g = dI_1/dV_{1/3}$ of the junction, with bias $V_{1/3}$ applied on the $v = 1/3$ side and current $I_1$ measured on the $v = 1$ side (for detailed results, see Appendix C). Current noise $S_{1/3}$ is also measured to evaluate the upstream heat transport along the 1/3-1 channels. Note that the same results are obtained from measurements in different configurations, e.g., with a bias applied on the $\nu = 1$ side (see Appendix D). We measured several junctions with different $L_g$ ranging from 0.3 to 40 μm. Here we mainly present the results for an 0.6-μm junction.

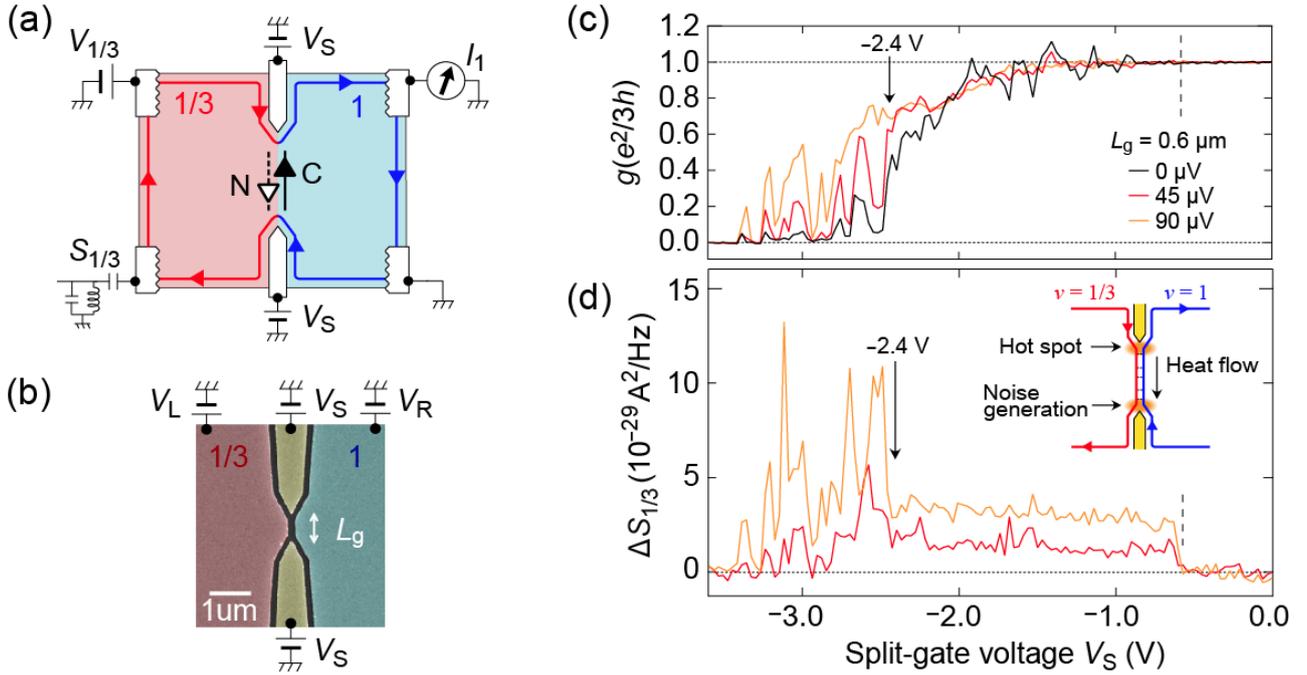

FIG. 2. (a) Schematic of experimental setup. The interacting 1/3-1 channels appear at the narrow junction, forming downstream charge (denoted by "C") and upstream neutral ("N") modes. We measured differential conductance $g = dI_1/dV_{1/3}$ and current noise $S_{1/3}$. (b) False-color scanning electron micrograph of $L_g = 0.6$ μm sample. The left (red) and right (blue) regions are set at $\nu = 1/3$ and $\nu = 1$, respectively, while the split gate (yellow) is energized to form a narrow 1/3-1 junction. (c) $V_S$ dependence of $g$ and (d) $\Delta S_{1/3}$ at several $V_{1/3}$ values of $L_g = 0.6$ μm sample. A narrow junction is formed below $V_S = -0.55$ V (indicated by dashed black lines). Zero-bias conductance oscillations and almost $V_S$-independent $\Delta S_{1/3}$ at $-2.4$ V $< V_S < -0.7$ V are signatures of coherent-incoherent crossover. (Inset) Schematic of noise generation process. Mixing of $\nu = 1/3$ and 1 channels with different chemical potentials causes heating at the hot spot, leading to the heat flow toward the noise-generating spot.

Figures 2(c) and (d), respectively, present the $V_S$ dependence of $g$ and bias-induced excess noise $\Delta S_{1/3}(V_{1/3}) \equiv S_{1/3}(V_{1/3}) - S_{1/3}(0)$, measured at several $V_{1/3}$ values. The vertical dashed lines mark $V_S \cong -0.55$ V, at which a narrow junction is formed. When $V_S$ is decreased slightly below $-0.55$ V, $g$ stays nearly unchanged at $G_e/3$, indicating that the impinging current is almost entirely transmitted across the junction. In terms of the 1/3-1 channel conductance, $g = G_e/3$ represents charge transport with conductance $2G_e/3$ from the downside to the upside. Notably, the formation of the narrow junction is signaled in $\Delta S_{1/3}$ as a jump from zero to a finite value. This signal indicates the existence of an upstream heat flow as there is no net charge flow through the back-reflected channel.

While the above behavior immediately below $V_S \cong -0.55$ V is nontrivial per se, it can be understood within the incoherent-transport picture. The signatures of the coherent-incoherent crossover emerge when $V_S$ further decreases to narrow the junction. At $-2.4$ V $< V_S < -0.7$ V, zero-bias conductance oscillates, representing disorder-specific electrical conductance. In contrast, $\Delta S_{1/3}$ is almost independent of $V_S$, indicating that the noise temperature of the junction is insensitive to $L$, the disorder configuration, and the electrical-conductance fluctuations. As detailed in the next section, $\Delta S_{1/3}$ measured at $V_{1/3} = 45$ μV, where the conductance oscillations are small but still visible, is close to the value expected for the quantized thermal conductance $G_Q/2$ of the single 1/3-1 channels [half of a conventional $v = 2/3$ device with top and bottom edges [3], see Figs. 1(e) and (f)]. All these observations correspond to the predicted crossover behaviors in the mesoscopic regime.

When $V_S$ is decreased below $V_S \cong -2.4$ V (highlighted by black arrows), zero-bias conductance rapidly decreases, indicating the breakdown of condition (1) by $L < L_T$ [3]. At finite bias, $\Delta S_{1/3}$ oscillates synchronized with $g$ down to $V_S \cong -3.4$ V, which is close to the pinch-off voltage ($-3.6$ V) of the junction at $B = 0$ T. In this regime, bias-induced tunneling through discrete levels dominates the transport [40], causing electron partitioning between the 1/3-1 channels and resultantly generating $\Delta S_{1/3}$ [41] (see also Appendix E). Below $V_S \cong -3.4$ V, $g = 0$ and $\Delta S_{1/3} = 0$ over the entire bias range, which is a result of the full decoupling the 1/3-1 channels caused as the disorder is no longer in action ($L < l$).

C. Dissipationless transport in the mesoscopic regime

We show that the weak $V_S$ dependence of $\Delta S_{1/3}$ [Fig. 2(d)] is linked with the quantized thermal conductance $G_Q/2$ predicted for the mesoscopic regime. To this end, we extend the theory of the incoherent transport regime [41, 42] that relates the thermal conductance of the 1/3-1 channels with the low-frequency current noise to the mesoscopic regime.

The mixing of the incoming channels with different chemical potentials causes heating at the endpoint of the downstream charge-mode flow ["hot spot"; see inset in Fig. 2(d)] and induces heat flow toward the other end, the endpoint of the upstream neutral-mode flow ("noise-generating spot"). The excess noise $\Delta S_{1/3}$ (and its counterpart $\Delta S_1$ on the $\nu = 1$ side) is generated at the noise-generating spot by dividing thermally excited electron-hole pairs into the outgoing $\nu = 1/3$ channel and the $\nu = 1$ channel that turns back to the hot spot and then exits from there [41]. This noise-generation mechanism naturally explains $\Delta S_{1/3} = \Delta S_1$ required by current conservation (see Appendix D). In the $L > L_{\text{in}}$ incoherent regime, the heat flow is diffusive due to inelastic processes, resulting in a gradual decrease of the effective temperature of the 1/3-1 channels from the hot spot (temperature $T_{\text{h}}$) toward the noise-generating spot ($T_{\text{ng}}$). Theory predicts $T_{\text{ng}} \propto 1/\sqrt{L}$ and hence $\Delta S_{1/3} = \Delta S_1 \propto 1/\sqrt{L}$ [41, 42]. Note that, unlike the current noise, which only reflects $T_{\text{ng}}$, the heat flows outgoing from the hot and noise-generating spots reflect $T_{\text{h}}$ and $T_{\text{ng}}$, respectively, and differ. The total outgoing power is determined by their sum and the energy dissipation during diffusive heat transport.

We use the same argument for the $L_{\text{T}} < L < L_{\text{in}}$ mesoscopic regime. Now, the 1/3-1 channels are short enough to warrant dissipationless charge- and neutral-mode transport, which leads to a uniform effective temperature $T_{\text{eff}}$ ($= T_{\text{h}} = T_{\text{ng}}$) established along the channels. Consequently, the heat generated at the hot spot splits equally between the two outgoing channels, meaning that the 1/3-1 junction serves as a 50:50 splitter, equipartitioning the injected heat into the two outgoing channels. As each of the incoming and outgoing channels has thermal conductance $G_{\text{Q}}$, it follows that the 1/3-1 junction has the quantized thermal conductance $G_{\text{Q}}/2$ in this case [3]. In the following, we calculate $T_{\text{eff}}$ by considering the balance between the input Joule power and the output heat flow [43] and compare it with the measured noise temperature $T_{\text{N}}$ defined as [41]

$$\Delta S_{1/3} \equiv 2k_{\text{B}}(T_{\text{N}} - T_0)G. \qquad (2)$$

If $T_{\text{N}} = T_{\text{eff}}$, the charge- and neutral-mode transport is dissipationless and therefore the 1/3-1 channels hold the thermal conductance $G_{\text{Q}}/2$ predicted for the mesoscopic regime.

The DC power supplied to the junction is given by $P_{\text{in}}^{\text{DC}} = V_{1/3}I_{\text{in}}/2 = G_{\text{e}}V_{1/3}^2/6$, where $I_{\text{in}} = G_{\text{e}}V_{1/3}/3$ is the impinging current. Similarly, the output DC power $P_{\text{out}}^{\text{DC}}$ is calculated as $P_{\text{out}}^{\text{DC}} = (V_{1/3}^{\text{out}}I_{1/3}^{\text{out}} + V_1^{\text{out}}I_1^{\text{out}})/2 = G_{\text{e}}V_{1/3}^2/6 - G(1 - 2G/G_{\text{e}})V_{1/3}^2$, where $V_{1/3(1)}^{\text{out}}$ and $I_{1/3(1)}^{\text{out}}$ are, respectively, the voltage and current of the outgoing $\nu = 1/3$ (1) channel and $G = I_1/V_{1/3}$ is the junction conductance. Thus, when the input Joule power,

$$\Delta P \equiv P_{\text{in}}^{\text{DC}} - P_{\text{out}}^{\text{DC}} = G(1 - 2G/G_{\text{e}})V_{1/3}^2, \qquad (3)$$

is balanced with the outgoing heat flow

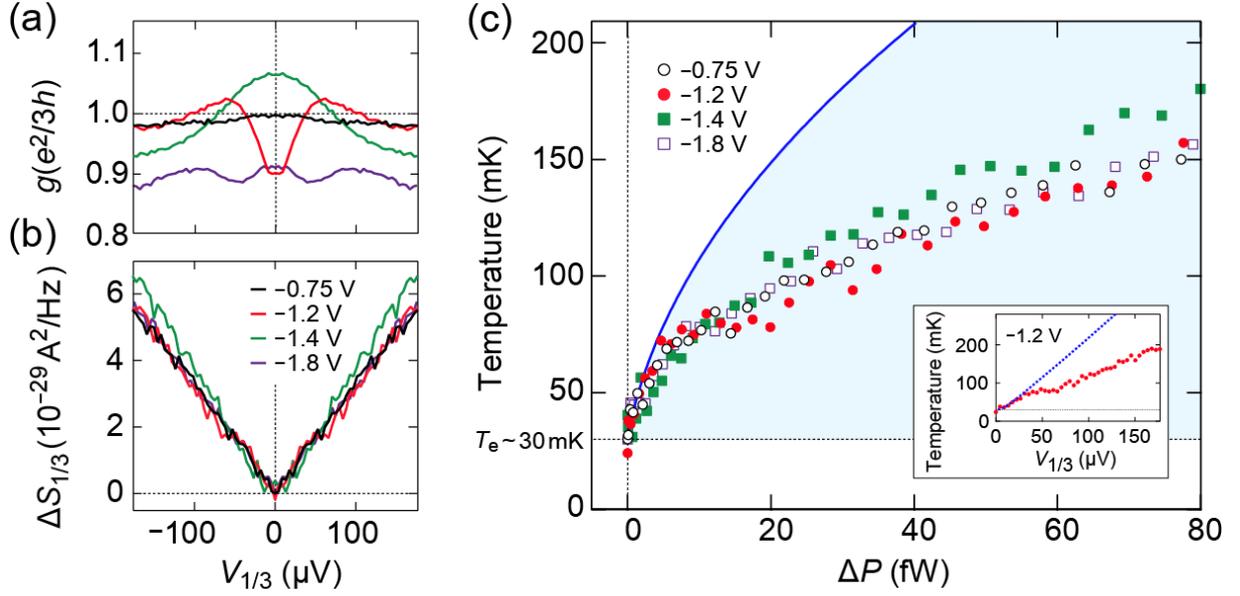

FIG. 3. (a) $V_{1/3}$ dependence of $g$ and (b) $\Delta S_{1/3}$ of the 0.6-μm sample measured at several $V_S$ values. Change in $V_S$ significantly varies the conductance behavior, but varies $\Delta S_{1/3}$ only slightly. The data are symmetrized against $V_{1/3}$ to improve the measurement resolution and compensate for minor self-gating effects (see Appendix D). (c) Noise temperature $T_N$ estimated from $\Delta S_{1/3}$ as a function of the input Joule power $\Delta P$. Solid blue line is the $T_{\text{eff}}$ curve obtained from eq. (5). The light-blue region indicates the possible $T_N$ range $T_e \leq T_N \leq T_{\text{eff}}$. The data approach the $T_{\text{eff}}$ curve at low bias, demonstrating dissipationless charge- and neutral-mode transport and quantized thermal conductance $G_Q/2$ of the 1/3-1 channels in the mesoscopic regime. (Inset) $V_{1/3}$ dependence of $T_N$ at $V_S = -1.2$ V. The data depart from the $T_{\text{eff}}$ curve (blue dotted line) above $V_{1/3} \cong 50$ μV, signaling the dissipation due to bias-induced inelastic processes at high bias. Results for wider junctions are available in Appendix F.

$$2J_Q^e = \pi^2 k_B^2 (T_{\text{eff}}^2 - T_e^2)/3h, \qquad (4)$$

where $J_Q^e$ is the heat flow through each outgoing channel, we have

$$T_{\text{eff}} = \sqrt{\frac{3h}{\pi^2 k_B^2}\Delta P + T_e^2}. \qquad (5)$$

Figure 3 shows the $V_{1/3}$ dependence of $g$ (a) and $\Delta S_{1/3}$ (b) at several $V_S$ values. While $g$ depends on $V_S$, all $\Delta S_{1/3}$ data exhibit a monotonic increase with $|V_{1/3}|$. The $\Delta S_{1/3}$ data for each $V_S$ are converted into $T_N$ and plotted in Fig. 3(c) against $\Delta P$ and compared with $T_{\text{eff}}$ obtained from eq. (5) (solid blue curve), where $\Delta P$ is calculated using eq. (3). The $T_N$ data fall in the region given by $T_e \cong 30$ mK $\leq T_N \leq T_{\text{eff}}$ (light blue region). Remarkably, below $\Delta P \cong 10$ fW, $T_N$ coincides with the $T_{\text{eff}}$ curve independently of the $V_S$ values, indicating the quantized thermal conductance $G_Q/2$. The departure of $T_N$ from $T_{eff}$ at higher $\Delta P$, in turn, signals the onset of bias-induced inelastic processes [44]. As one would expect, the onset voltage $|V_{1/3}| \cong 50$ μV (inset) of the departure is close to the bias above which the electrical-conductance fluctuations disappear. We add that the $\Delta P$ dependence of $T_N$ at high bias suggests heat transport through the 2D bulk as the dominant contributor to dissipation (see Appendix G).

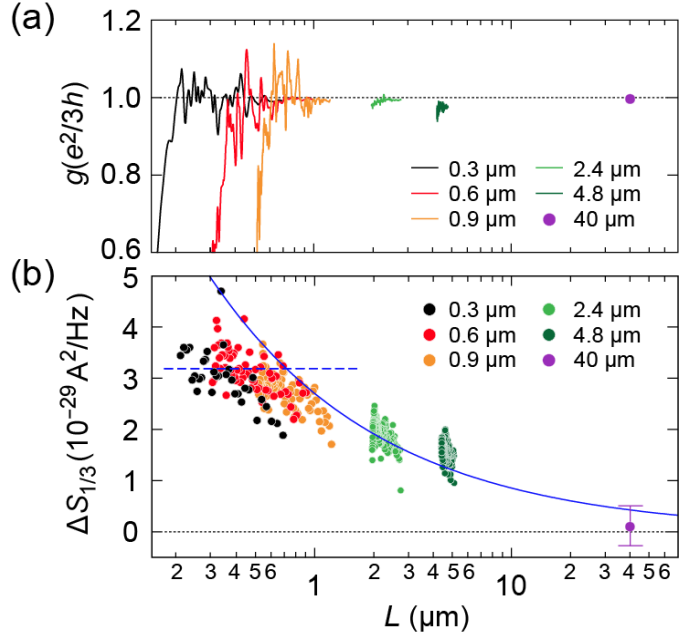

FIG. 4. (a) Zero-bias conductance and (b) $\Delta S_{1/3}$ at $V_{1/3} = 90$ μV as a function of the 1/3-1 channel length $L$ for several samples with different $L_g$. The $g$ and $\Delta S_{1/3}$ data for the 0.6-μm sample are the same as those in Fig. 2. Narrow junctions ($L < 1$ μm) show the signatures of coherent-incoherent crossover, namely the mesoscopic electrical-conductance fluctuations and almost constant $\Delta S_{1/3}$. The blue dashed line in (b) is a visual guide for constant $\Delta S_{1/3}$. Wide junctions ($L > 1$ μm) show the suppression of electrical-conductance fluctuations and the decrease in $\Delta S_{1/3}$ with increasing $L$, indicating the inelastic processes in the junction. The blue solid line is a visual guide $\Delta S_{1/3} \propto 1/\sqrt{L}$ for the length dependence of $\Delta S_{1/3}$ in the incoherent regime [41]. The crossover between the narrow and wide junction regions suggests $L_{in} \sim 1$ μm inelastic scattering length.

### D. Inelastic scattering length

We performed similar measurements on several 1/3-1 junctions with different $L_g$. Figure 4 plots the results as a function of $L$ estimated from $V_S$ using a 3D finite element method for each $L_g$ (see Appendix H). Above $L \sim 1$ μm, the electrical-conductance fluctuations are suppressed, and $\Delta S_{1/3}$ decreases monotonically with increasing $L$, suggesting $L_{in}$ ~ 1 μm in our devices (at $T_e$ = 30 mK).

The onset lengths for the suppression of electrical-conductance fluctuations and the decline of thermal conductance, referred to as the charge and heat equilibration lengths, respectively, can differ when the bare interaction is very strong ($\Delta \cong 1$) [16-21]. Conversely, when the bare interaction is not too strong as in the present case ($\Delta \cong 1.2$), these equilibration lengths become close to each other. Indeed, our results show that this is the case and that both equilibrium lengths can be represented by a single inelastic scattering length $L_{in}$ ~ 1 μm, as assumed in Ref. [3].

Once $L_{in}$ is known, one can estimate the neutral-mode velocity $v_\sigma \sim k_B T_e L_T / h$, a measure of the renormalized interaction strength $\Delta_r$. Theory predicts that a rough relationship between $L_{in}$ and other characteristic lengths, including $L_T$,

$$L_{in} \sim \frac{1}{\Delta - 1} \frac{L_T^2}{l} \qquad (6)$$

holds at $L_T > l$ [3]. With $L_{in}$ ~ 1 μm, $\Delta \cong 1.2$, and $l$ ~ 30 nm (see Appendix I), eq. (6) gives the thermal length $L_T$ ~ 80 nm and $v_\sigma$ ~ 50 m/s at $T_e$ = 30 mK. This $v_\sigma$ value is significantly smaller than the mode velocities, 24 and 155 km/s, expected for the bare interaction strength $\Delta \cong 1.2$ (see Appendix B). The significant reduction of $v_\sigma$ implies $\Delta_r \cong 1$, namely the renormalization of the inter-channel interaction to the vicinity of the strong-disorder fixed point (for further discussions, see Appendix I).

## III. SUMMARY

We have demonstrated the coherent-incoherent crossover of the charge- and neutral-mode transport in interacting 1/3-1 channels that replicate the $v$ = 2/3 edge channel. This was made possible by virtue of the in-situ tuning of $L$ even below the $L_{in}$ ~ 1 μm characteristic of our 1/3-1 junction. We observed fluctuating electrical conductance and current noise that corresponds to the quantized thermal conductance, the signatures of the coherent-incoherent crossover. In this mesoscopic regime, the electrical conductance exceeds $g = G_e/3$ at some $V_S$ values [see Fig. 2(c)], signaling the Andreev-like reflection of fractional quasiparticles at the 1/3-1 junction [40, 45-47]. This observation highlights the Andreev-like reflection as the driving force for the renormalization toward the strong-disorder fixed point. This study demonstrates a junction of topologically distinct systems as an exquisite platform for probing electron dynamics in interacting edge channels and, in a broader sense, for examining fundamental concepts in condensed matter physics, such as particle-hole symmetry and elementary excitations in correlated systems.

## ACKNOWLEDGMENTS

The authors thank T. Jonckheere, J. Rech, T. Martin, and T. Fujisawa for fruitful discussions and H. Murofushi and M. Imai for technical support. This work was supported by Grants-in-Aid for Scientific Research (Grant nos. JP16H06009, JP19H05603, and JP22H00112) and JST PRESTO Grant no. JP17940407.

M.H. conceived this study. M.H., T.A., and S.S. fabricated the sample. M.H. performed the measurement and analysis. T.I. and N.S. performed the DMRG calculations. M.H., T.I., N.S., and K.M. interpreted the results. M.H. wrote the paper with help from N.K., N.S., and K.M. All authors discussed the results and commented on the paper.

## APPENDIX A: DMRG CALCULATION

We calculated the ground state wave function of 2DESs and the $n_e$ profiles at the 2/3 edge and 1/3-1 junction using the DMRG calculation method for a torus geometry [36, 48-51]. We assumed that electron spin is fully polarized. The system size is defined by the lengths $L_x$ and $L_y$ of the unit cell and relates to the number of magnetic flux quanta $M$ = 210 as $L_x L_y = 2\pi l_B^2 M$. We set the aspect ratio of the system at $L_x/L_y = 0.5$. The electrostatic potential is induced by modulating the background-positive-charge density $n_p(x)$ in the doped layer, located above the 2DES with vertical distance $d$ [Figs. 5(a) and (b)]. For the 1/3-1 junction (2/3 edge), the $n_p(x)$ profile has two regions with the ratio $n_p/n_\varphi$ = 1/3 (2/3) and $n_p/n_\varphi = 1$ (0), and the two regions are connected with the abrupt change in $n_p/n_\varphi$ at their boundary. Because of the finite distance $d$, the electrostatic potential induced in the 2DES layer gently varies with $x$. We adjusted the total electron number in the system to form the two well-developed QH states in the bulk regions. The DMRG calculation result for the full-range $n_e(x)$ profile of the 1/3-1 junction, obtained with $d/L_x = 0.002$, is shown in Fig. 5(c). The data shown in Figs. 1(c) and (d) were also obtained with the same $d/L_x$ value. Even when the total electron number, the $n_p(x)$ profile, or $d/L_x$ slightly varies, we found similar $n_e$ profiles with the signatures of the counter-propagating 1/3-1 channels, as long as $d/L_x$ is small enough to form well-developed bulk QH regions. Further technical details are available in Ref. [36].

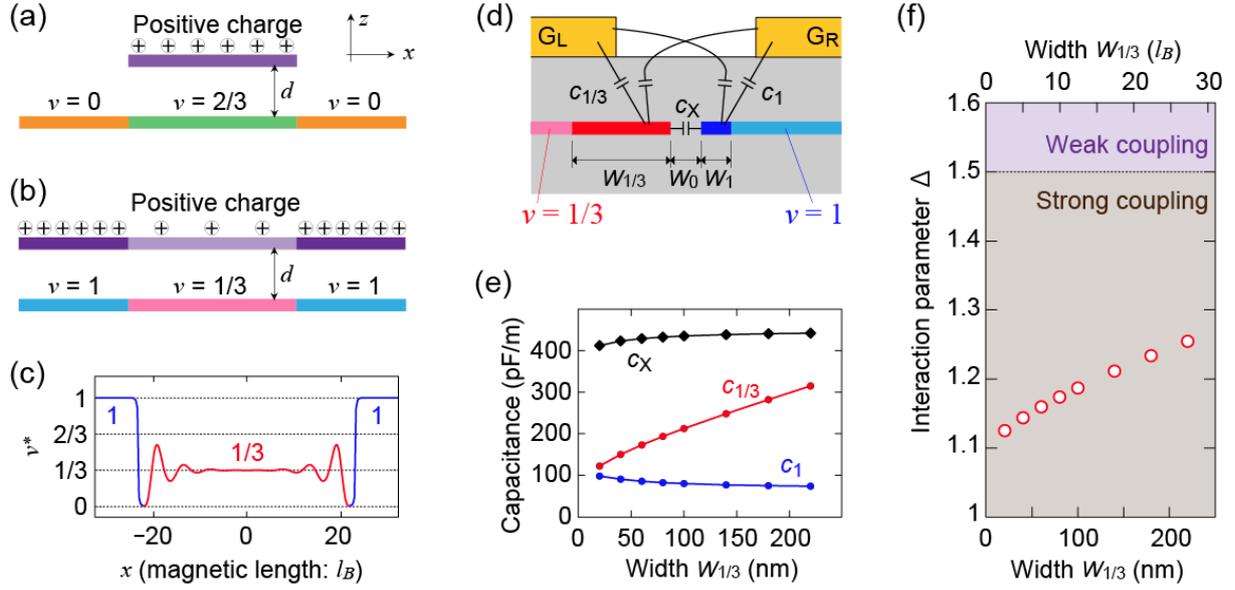

FIG. 5. (a) Schematic of the positive-charge configurations used for DMRG calculations of 2/3-0 and (b) 1/3-1 junctions. Positive background charges (density $n_p$) locate at a vertical distance $d$ from the 2DES. Abrupt changes in $n_p$ induce the electrostatic potentials in the 2DESs to form the QH junctions, gently varying near the junctions. (c) Full-range $v^*$ profile of the 1/3-1 junction obtained by the DMRG calculation. (d) Schematic cross-section of the 1/3-1 junction with capacitive couplings. Red and blue regions respectively depict the compressible strips of the $v = 1/3$ and $v = 1$ channels, while pink and cyan regions are the incompressible bulk regions. Yellow regions denoted as $G_L$ and $G_R$ are left and right surface-gate electrodes, respectively. (e) Distributed channel capacitances estimated by a 2D finite element method. While the DMRG calculation result shows $W_1 \cong 10.5$ nm and $W_0 \cong 13$ nm, the oscillations of $v^*$ in the $v = 1/3$ region makes it difficult to estimate $W_{1/3}$. Therefore, we calculated the capacitances for several $W_{1/3}$ values. (f) Bare inter-channel interaction parameter $\Delta$ estimated using the capacitance values in (e). The $\Delta$ values stay at $\Delta \cong 1.2$ over the entire range of $W_{1/3}$ used for the calculation, indicating that the 1/3-1 channels are in the strong-coupling phase.

## APPENDIX B: DISTRIBUTED CAPACITANCE MODEL

We numerically estimate $\Delta$ using the distributed capacitance model of Coulomb interaction, which has been used to investigate charge dynamics along QH edge channels [15, 31, 37, 38, 52-54]. The $n_e(x)$ profile in Fig. 1(d) enables us to estimate widths $W_{1/3}$ and $W_1$ of compressible strips of the $v = 1/3$ and $v = 1$ channels, respectively, and that of the $v^* = 0$ depletion region $W_0$. While the $n_e(x)$ profile directly tells us that $W_1 \cong 1.3 l_B \cong 10.5$ nm and $W_0 \cong 1.6 l_B \cong 13$ nm (the magnetic length $l_B = 8.1$ nm at $B = 10$ T), oscillations of $v^*$ in the $v = 1/3$ region make the estimation of

$W_{1/3}$ difficult. Therefore, we performed the following calculations for several values of $W_{1/3}$ in 20 nm $\leq W_{1/3} \leq$ 220 nm. Figure 5(d) shows a schematic cross-section of the 1/3-1 junction with the geometric coupling capacitances $c_\alpha$ ($\alpha$ = 1/3 or 1) and $c_X$, where $c_\alpha$ is between the compressible $\nu = \alpha$ strip and the gate metals and $c_X$ is between the compressible strips. Figure 5(e) plots these capacitances computed using a 2D finite element method with commercial software COMSOL. We assumed the incompressible bulk QH regions as insulators in this calculation. Because the compressible strips locate close to each other, $c_X$ is larger than $c_{1/3}$ and $c_1$ over the entire range of $W_{1/3}$ we assumed.

The interaction strength $\Delta$ is related to the velocity parameters $v_1$, $v_{1/3}$, and $v_X$ as $\Delta \equiv (2 - \sqrt{3}p)/\sqrt{1-p^2}$, where $p \equiv (2v_X/\sqrt{3})/(v_1 + v_{1/3})$. Here, $v_1$ and $v_{1/3}$ respectively reflect the interactions within $\nu = 1$ and $\nu = 1/3$ channels, and $v_X$ reflects the repulsive inter-channel interaction [1, 2]. When we ignore minor contributions of the electrochemical capacitance of the edge channels, these velocity parameters are expressed using the geometric coupling capacitances as [31]

$$v_1 = \frac{e^2}{h} \frac{c_{1/3} + c_X}{c_1 c_{1/3} + (c_1 + c_{1/3})c_X}, \quad (7)$$

$$v_{1/3} = \frac{e^2}{3h} \frac{c_1 + c_X}{c_1 c_{1/3} + (c_1 + c_{1/3})c_X}, \quad (8)$$

$$v_X = \frac{e^2}{h} \frac{c_X}{c_1 c_{1/3} + (c_1 + c_{1/3})c_X}. \quad (9)$$

By substituting the capacitance values estimated above, we find $\Delta \cong 1.2$ over the entire range of $W_{1/3}$ [Fig. 5(f)], which ensures that the 1/3-1 channels are in the strong-coupling phase.

The distributed capacitance model enables us to obtain the speeds of the counter-propagating eigenmodes expected for the bare interaction strength ($\Delta \cong 1.2$), 24 km/s and 155 km/s, by solving the wave equation

$$\frac{\partial}{\partial t}\begin{pmatrix}\rho_1\\\rho_{1/3}\end{pmatrix} = \begin{pmatrix} v_1 & v_X \\ -v_X & -v_{1/3} \end{pmatrix} \frac{\partial}{\partial x}\begin{pmatrix}\rho_1\\\rho_{1/3}\end{pmatrix}, \quad (10)$$

where $\rho_1$ and $\rho_{1/3}$ are the charge in the $\nu = 1$ and 1/3 channels at position $x$ and time $t$ [15].

**APPENDIX C: EXPERIMENTAL SETUP**

**Sample Preparation**

The 1/3-1 junctions were fabricated in an 80-μm-wide Hall bar with the 2DES confined to a 30-nm-wide GaAs quantum well (QW). The center of the QW is located 190 nm below the surface. The heterostructure has a 65-nm-thick

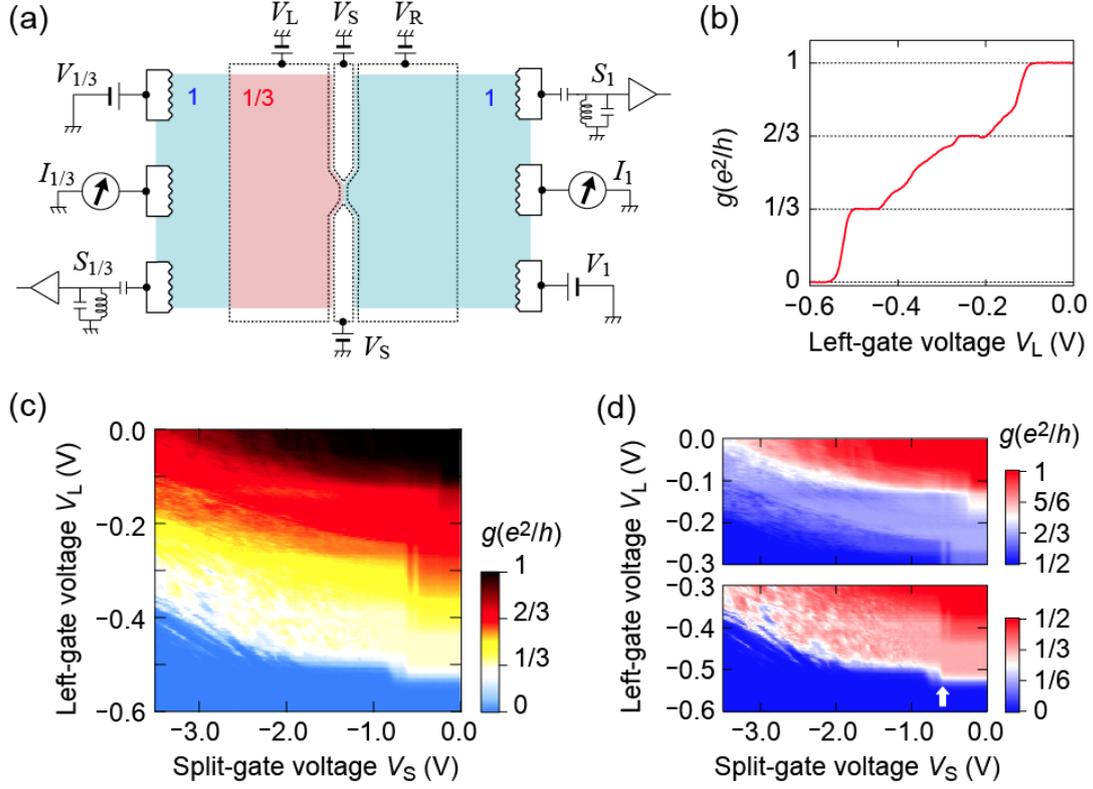

FIG. 6. (a) Schematic of the whole device and measurement setup. Dotted lines indicate the geometries of surface-gate electrodes. The 2DESs below the right surface gate ($V_R = 0$ V) and outside the gated area are in the $\nu = 1$ state (blue regions) at $B = 10$ T. The $\nu = 1/3$ state (red region) is formed by decreasing the electron density below the left surface gate with $V_L \cong -0.45$ V [see data in (b)]. The long 1/3-1 channels across the 80-μm-wide Hall bar, located immediately to the left of the $\nu = 1/3$ region, are fully equilibrated to produce a unidirectional channel of electrical conductance $2e^2/3h$. In the standard Landauer-Büttiker edge transport picture, the setup can be expressed simply as that in Fig. 2(a) [40]. (b) $V_L$ dependence of zero-bias conductance $g$ at $V_R = V_S = 0$ V. (c) Color plot of $g$ as a function of $V_S$ and $V_L$ at $V_R = 0$ V. (d) Same $g$ data as (c) plotted with different ranges of $g$ and $V_L$: (upper panel) $e^2/2h < g < e^2/h$ and $-0.3$ V $< V_L < 0$ V, (lower panel) $0 < g < e^2/2h$ and $-0.6$ V $< V_L < -0.3$ V. The conductance oscillations signaling the mesoscopic electrical-conductance fluctuations are observed for the 1/3-1 junction at $V_L \cong -0.45$ V and not for the 2/3-1 junction at $V_L \cong -0.2$ V. The kink structure of the $g \cong e^2/3h$ region at $V_S \cong -0.55$ V, highlighted by the white arrow in the lower panel, indicates the formation of the narrow 1/3-1 junction.

$Al_{0.33}Ga_{0.67}As$ spacer layer between the QW and the doped layer. The $n^+$-GaAs substrate serves as a back gate, and metal surface electrodes of 30-nm-thick gold on 10-nm-thick titanium work as surface gates. We patterned the sample using e-beam lithography for fine surface-gate structures and photolithography for in-plane semiconductor structures, Au-Ge-Ni alloyed ohmic contacts, and coarse metalized structures.

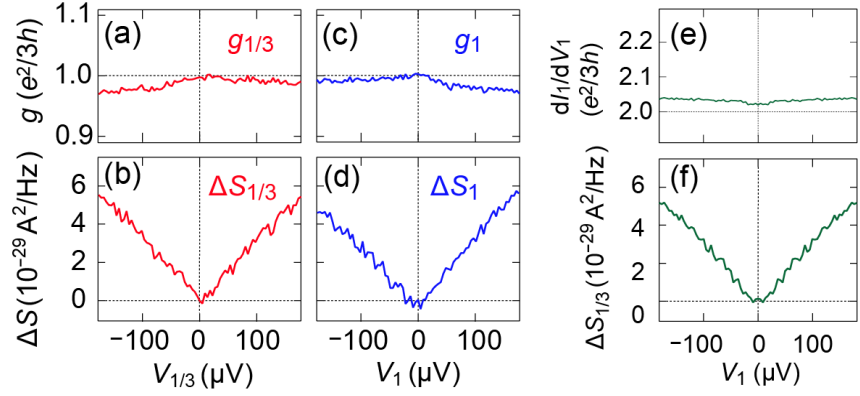

FIG. 7. (a) $V_{1/3}$ dependence of $g_{1/3}$ and (b) $\Delta S_{1/3}$ of the 0.6-μm sample measured at $V_S = -0.75$ V. (c) $V_1$ dependence of $g_1$ and (d) $\Delta S_1$. (e) Symmetrized $dI_1/dV_1$ at $V_S = -0.75$ V, measured by applying $V_1$. Back reflection from the 1/3-1 junction gives $dI_1/dV_1 \cong 2e^2/3h$, signaling the charge-mode transport with conductance $2e^2/3h$ from the downside to the upside of the junction. (f) Symmetrized $\Delta S_{1/3}$ data measured simultaneously with the data in (e).

**Measurement setup.**

Figure 6(a) shows the complete schematic of the experimental setup. The samples were placed in a dilution refrigerator of base temperature $T_{base} = 8$ mK. The electron temperature $T_e = 30$ mK was estimated by measuring the thermal noise of a 1/3-1 junction device. We applied a magnetic field of $B = 10$ T perpendicular to the 2DES to form the 1/3-1 junction. The electron density of the 2DES in the $\nu = 1$ region (blue region) is set at $2.4 \times 10^{11}$ cm$^{-2}$ by applying a back-gate voltage of 1.6 V, while that in the $\nu = 1/3$ region (red region) is at $0.8 \times 10^{11}$ cm$^{-2}$ by applying $V_L \cong -0.45$ V [see Figs. 6(b)-(d)]. For the main results, we measured dc transport properties using the standard lock-in technique with the 10-μV ac modulation of $V_{1/3}$ at 79 Hz. The current-noise characteristics were measured with inductor-capacitor resonance circuits and homemade-HEMT-based voltage amplifiers [the details of the measurement setup are available in Refs. [55, 56].

**APPENDIX D: RESULTS IN DIFFERENT SETUPS**

While we mainly described the measurement results for $I_1$ and $S_{1/3}$ with $V_{1/3}$ applied, we also performed other measurements in different experimental configurations [see Fig. 6(a)]. Different choices of the measurement quantities ($I_1$ or $I_{1/3}$ and $S_1$ or $S_{1/3}$) or the applied bias ($V_1$ or $V_{1/3}$) do not change our conclusion, and they demonstrate the reproducibility of the experimental results, as explained below for a few examples.

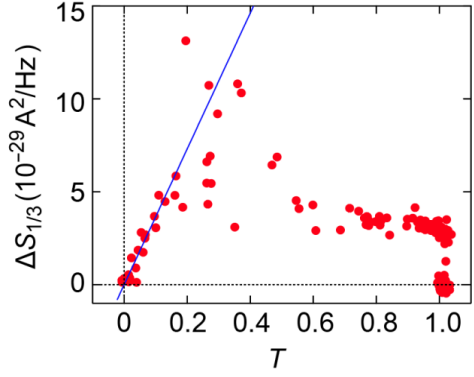

FIG. 8. Transmission probability dependence of $\Delta S_{1/3}$.

Figures 7(a) and (b), respectively, display the (non-symmetrized) raw data of the differential conductance $g_{1/3}$ = $dI_1/dV_{1/3}$ and $\Delta S_{1/3}$ of the 0.6-μm junction at $V_S$ = −0.75 V measured with $V_{1/3}$ applied. Figures 7(c) and (d) show their counterparts, $g_1 = dI_{1/3}/dV_1$ and $\Delta S_1$, respectively, measured with $V_1$ applied. Both the junction conductance and current noise show similar bias dependence independently of the choice of the ohmic contact to bias, $V_{1/3}$ or $V_1$, reflecting the Onsager-Casimir reciprocal relations for the two-terminal transport measurements. The weak asymmetric behaviors in these plots reflect the self-gating effect in the $\nu$ = 1/3 region by the finite source-drain bias. Because the self-gating effect is out of the scope of this study, we symmetrized the $g_{1/3}$ or $\Delta S_{1/3}$ data by averaging the values in opposite bias directions to compensate for the asymmetric behaviors and improve the signal-to-noise ratio [see Figs. 3(a) and (b)].

Figures 7(e) and (f) respectively show the symmetrized differential conductance $dI_1/dV_1$ and $\Delta S_{1/3}$ at $V_S$ = −0.75 V measured with $V_1$ applied. The measured $dI_1/dV_1$ is close to $2e^2/3h$, directly demonstrating that the 1/3-1 channels have electrical conductance $2e^2/3h$ from the downside to the upside of the junction. The current noise $\Delta S_{1/3}$ in (f) is close to that observed by varying $V_{1/3}$ [Fig. 3(b)]. Note that the $\Delta S_{1/3}$ values in (f) is similar to $\Delta S_1$ in (d), indicating current conservation $\Delta S_{1/3} = \Delta S_1$ in the present device.

## APPENDIX E: TRANSMISSION PROBABILITY DEPENDENCE OF $\Delta S_{1/3}$

Figure 8 shows the same $\Delta S_{1/3}$ data as those in Fig. 2(d) plotted as a function of the transmission probability $T$ = $I_1/I_{in}$. The experimental data near $T$ = 0 fit well with the shot-noise curve in the strong-backscattering limit calculated assuming $\Delta S_{1/3} = 2eI_1$ (blue solid line). This agreement indicates that $g$ and $\Delta S_{1/3}$ at $-3.4$ V $< V_S <-2.4$ V are well understood in terms of shot-noise generation due to electron partitioning processes between the $\nu = 1/3$ and 1 channels through discrete levels [40, 41].

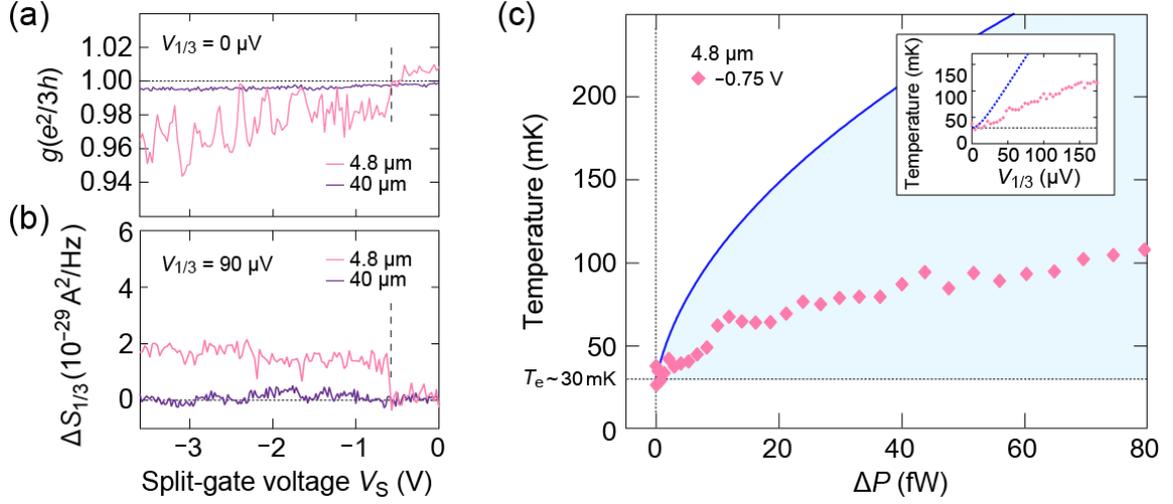

FIG. 9. (a) $V_S$ dependence of $g$ at $V_{1/3} = 0$ μV and (b) $\Delta S_{1/3}$ at $V_{1/3} = 90$ μV of 4.8- and 40-μm samples. (c) $T_N$ of the 4.8-μm sample plotted as a function of $\Delta P$, measured at $V_S = -0.75$ V. (Inset) $V_{1/3}$ dependence of $T_N$ at $V_S = -0.75$ V.

**APPENDIX F: RESULTS FOR WIDE JUNCTIONS**

Figure 9 summarizes the measurement results for wider 1/3-1 junctions. Below $V_S = -0.55$ V (indicated by dashed black lines), the 4.8-μm sample shows zero-bias electrical-conductance oscillations and finite $\Delta S_{1/3}$ almost independent to $V_S$, similarly to the 0.6-μm sample [see Figs. 2(c) and (d)]. However, the conductance oscillation amplitude and the $\Delta S_{1/3}$ value are smaller than those of the 0.6-μm sample, reflecting the 4.8-μm 1/3-1 channels being in the incoherent transport regime. The 40-μm sample shows no conductance fluctuations and $\Delta S_{1/3} \cong 0$, indicating that neutral mode is fully attenuated due to more influential inelastic processes in the longer channels. Figure 9(c) plots $T_N$ for the 4.8-μm sample estimated from the zero-bias conductance in (a) and the $\Delta S_{1/3}$ data in (b). The data are well below the $T_{\text{eff}}$ curve over the entire range of $\Delta P$, showing no approach to the curve at low bias. The data in (c) and the inset indicate the energy dissipation from the 4.8-μm 1/3-1 channels even at low bias.

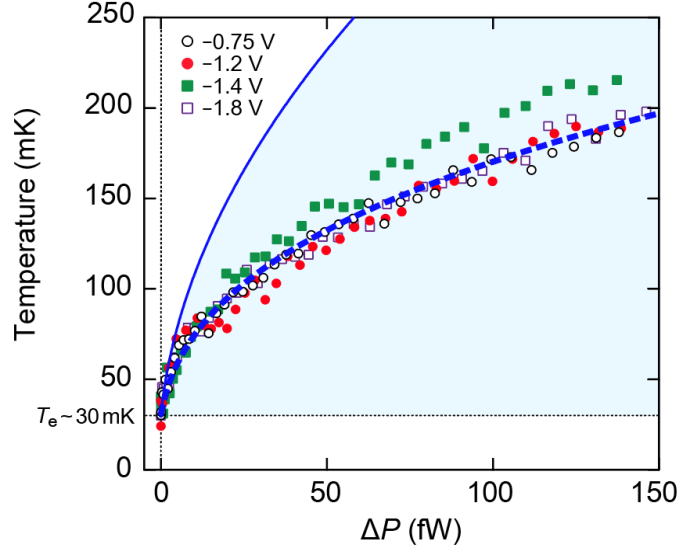

FIG. 10. $\Delta P$ dependence of $T_N$ measured for 0.6-μm junction at several $V_S$ values [the same data as in Fig. 3(c)] plotted with the $T_b$ curve obtained from eq. (11) with $\alpha = 15$ pW/K$^3$ (blue dashed line).

## APPENDIX G: ENERGY DISSIPATION AT HIGH BIAS

Inter-mode scattering causes the dissipation from the 1/3-1 channels in the incoherent regime. The heat transport through the 2D bulk of QH systems, which scales proportionally to the cube of temperature, may mediate the heat outflow from the 1/3-1 junction [3, 10, 11, 57].

In the main text, we calculated $T_{eff}$ of the 1/3-1 channels by considering only the heat transfer through the outgoing $\nu = 1/3$ and 1 channels and compared it with $T_N$ [Fig. 3(c)]. Whereas $T_N$ is close to $T_{eff}$ at low bias, $T_N$ becomes smaller than $T_{eff}$ at high bias, indicating energy dissipation due to inelastic inter-mode scattering. To examine the possible contribution for the dissipation by heat transport through the bulk, we calculate the effective temperature $T_b$ by taking the bulk-contribution term into account using the following equation,

$$\Delta P = \frac{\pi^2 k_B^2}{3h}(T_b^2 - T_e^2) + \alpha_b(T_b^3 - T_e^3), \qquad (11)$$

where $\alpha_b$ is a fit parameter. Figure 10 compares the measured $\Delta P$ dependence of $T_N$, the same data as those in Fig. 3(c), with $T_b$ simulated using eq. (11) with $\alpha_b = 15$ pW/K$^3$ (blue dashed line). The experimental data fit the $T_b$ curve, suggesting that the heat transport through the 2D bulk dominantly contributes to the dissipation at high bias. Note that, in contrast to previous experiments, where the electron-phonon coupling in a small diffusive metal promotes the phonon contribution [43], heat flow toward the cold phonon bath can be small in our device with no such diffusive metal.

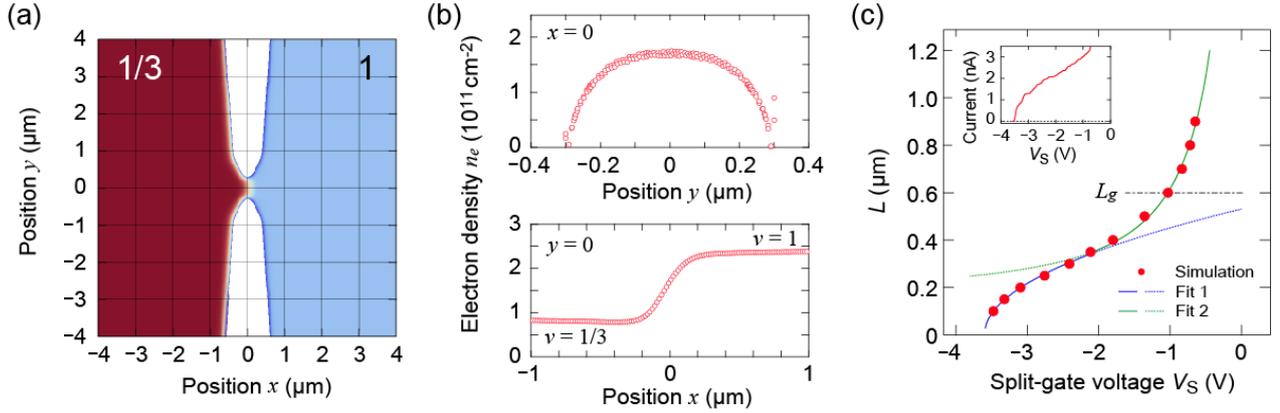

FIG. 11. (a) Example of 2DES shapes used for the finite element method calculation of the electron density profile ($V_S \cong -1$ V for the 0.6-μm sample). (b) Calculated electron-density profiles at $V_S = -1.03$ V along the $x = 0$ (upper) and $y = 0$ (lower) lines in (a). (c) Channel length $L$ plotted as a function of $V_S$. Red circles are data points obtained from the finite element method calculation. Green and blue curves are the fits using eqs. (12) and (13), respectively. (Inset) Pinch-off trace of the 0.6-μm sample at $B = 0$ T. The surface-gate and back-gate voltages are set at the same values as those to form the 1/3-1 junction at $B = 10$ T.

**APPENDIX H: ESTIMATION OF THE 1/3-1 CHANNEL LENGTH**

We estimated the 1/3-1 channel length $L$ for each $L_g$ sample from a zero-magnetic-field 2D electron-density profile near the junction, which we calculated using the following procedure. 1) We regarded the 2DES in the QW as a 30-nm-thick metallic plate and designed a plausible in-plane 2DES shape. 2) We calculated the electron density profile induced on the plate by the surface gate voltages using a three-dimensional (3D) finite element method. If the obtained electron-density profile at the 2DES edge is nearly zero, we consider the designed 2DES shape reasonable for the given surface gate voltages. Figure 11(a) shows an example of the 2DES shape obtained by this method. 3) We again calculated the electron density profile using the 2DES shape with fine-tuning of $V_S$ to find the correct $V_S$ value that gives exactly zero electron density at the 2DES edge near the constriction [see Fig. 11(b)]. By regarding the width of the constriction as $L$, we obtained a set of $V_S$ and $L$ corresponding to each other. We repeated the above procedure for different $L$ values and obtained several sets of ($L$, $V_S$). Figure 11(c) shows the results for the 0.6-μm sample, as an example. To attain the $V_S$ dependence of $L$ over the entire range of $V_S$, we fitted the data points of ($L$, $V_S$) using the phenomenological functions

$$L = \alpha + \frac{\beta}{(V_S - V_1)^2 + \gamma}, \quad (12)$$

$$L = \chi\sqrt{V_S - V_2}, \quad (13)$$

where α, β, γ, χ, $V_1$, and $V_2$ are fit parameters. The function (12) is used to fit the data at $L$ ≥ 0.3 μm, while (13) is used at $L$ ≤ 0.3 μm. With these functions, we converted $V_S$ to $L$ and plotted the data as functions of $L$ in Fig. 4.

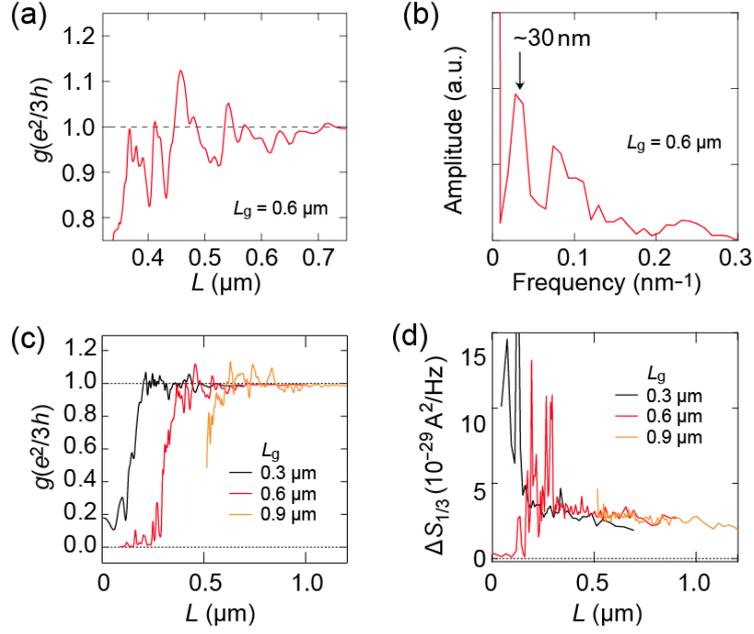

FIG. 12. (a) Magnified view of the zero-bias electrical-conductance fluctuations of the 0.6-μm sample shown in Fig. 4(a). (b) Fourier transform spectrum of the conductance fluctuations in (a) in 400 nm ≤ $L$ ≤ 740 nm. (c) $L$ dependence of the zero-bias conductance and (d) $\Delta S_{1/3}$ at $V_{1/3}$ = 90 μV of $L_g$ = 0.3, 0.6 and 0.9-μm junctions plotted on a linear

**APPENDIX I: CHARACTERISTIC LENGTHS**

Figure 12 (a) shows the magnified view of the zero-bias electrical-conductance fluctuations in Fig. 4(a). We estimated $l \cong 30$ nm from the oscillation period of the fluctuations that is presented as the peak in Fig. 12(b). The obtained $l \cong 30$ nm roughly corresponds to the thickness of the $Al_{0.33}Ga_{0.67}As$ spacer layer (65 nm) of the heterostructure.

One may consider that it is possible to estimate $L_T$ from the zero-bias pinch-off traces of the 1/3-1 junctions because the $v = 1/3$ and 1 channels are decoupled at $L < L_T$ for the direct current [3]. Figures 12(c) and (d) show the $L$ dependence of the zero-bias conductance and $\Delta S_{1/3}$ at $V_{1/3} = 90$ μV, respectively, of $L_g = 0.3$, 0.6, and 0.9-μm junctions, plotted on a linear scale of $L$. Interestingly, these devices show pinch-offs at different lengths (0.15, 0.3, and 0.5 μm for the $L_g = 0.3$, 0.6, and 0.9-μm junctions, respectively). The junctions show current-noise generation due to bias-induced electron partitioning below these pinch-off lengths, as shown in (d), corresponding to the conductance behaviors. These observations suggest that the $L_T$ values are different from each other for these junctions. We consider that the difference is caused by the variations of the influence from the negative $V_S$. The larger $L_g$ device shows a pinch-off at lower $V_S$, and the lower $V_S$ promotes the decoupling of the $v = 1/3$ and 1 channels. In the main text, we estimated $L_T$ using eq. (6), instead of the pinch-off length, to exclude the complication that may come into play due to the influence of $V_S$. We note that, however, our conclusion does not change even if we estimate $L_T$ from the pinch-off traces presented here.